\documentclass[universe,article,accept,pdftex,moreauthors]{Definitions/mdpi} 
\firstpage{1} 
\pubvolume{0}
\issuenum{0}
\articlenumber{0}
\pubyear{2024}
\copyrightyear{2024}
\datereceived{2024 April 26}
\daterevised{2024 May 29} 
\dateaccepted{2024 } 
\datepublished{2024  } 
\hreflink{https://doi.org/} 

\usepackage{natbib}

\newcommand{\kms}{km\,s$^{-1}$}

\Title{On the nature of the radio calibrator and gamma-ray emitting 
NLS1 galaxy 3C 286 and its multiwavelength variability }

\TitleCitation{3C 286}

 \Author{S. Komossa$^{1}$*, S. Yao$^{2,1}$, D. Grupe$^{3}$, A. Kraus$^{1}$}

\AuthorNames{S. Komossa, et al.}

\AuthorCitation{S. Komossa, et al.}

\address{%
$^{1}$ \quad Max-Planck Institut f\"ur Radioastronomie, Auf dem H\"ugel 69, 53121 Bonn, Germany\\
$^{2}$ \quad 
National Astronomical Observatories, Chinese Academy of Science, Beijing 100101, China\\
$^{3}$ \quad Department of Physics, Geology, and Engineering Technology, Northern Kentucky University, 1 Nunn Drive, Highland Heights, KY      41099, USA\\
}

\corres{Correspondence: skomossa@mpifr.de (S.K.)
}

\abstract{The quasar 3C 286, a well-known calibrator source in radio astronomy, was found to exhibit exceptional multiwavelength properties.
Its rich and complex optical emission-line spectrum revealed its narrow-line Seyfert 1 (NLS1) nature.
Given its strong radio emission, this makes 3C 286 one of the radio-loudest NLS1 galaxies known to date. 3C 286
is also one of very few known compact steep-spectrum (CSS) sources detected in the gamma-ray regime.
Observations in the X-ray regime, rarely carried out so far, revealed evidence for variability,
raising the question if driven by the accretion disk or jet. 3C 286 is also well known
for its damped Lyman alpha system from an intervening absorber at $z$ = 0.692,
triggering a search for the corresponding X-ray absorption along the line-of-sight.
Here, we present new observations in the radio, X-ray, optical and UV band.
The nature of the X-ray variability is addressed. Spectral evidence suggests that it is primarily driven by the accretion disk (not the jet), and the X-ray spectrum is well fit by a powerlaw plus soft excess model. The radio flux density and polarization remain constant at the Effelsberg telescope resolution, reconfirming the use of 3C 286 as radio calibrator.  
The amount of reddening/absorption along the line-of-sight {\em{intrinsic}} to 3C 286 is rigorously assessed. None is found, validating the derivation of a high Eddington ratio ($L/L_{\rm Edd} \sim 1$) and of the very high radio-loudness index of 3C 286. Based on the first deep Chandra  image of 3C 286, tentative evidence for hard X-ray emission from the SW radio lobe is reported. 
A large variety of models for the gamma-ray emission of 3C 286 is briefly
discussed. 
}


\begin{document}


\section{Introduction: 3C 286}

The quasar 3C 286 serves as one of the most important calibrator sources in  the centimeter radio bands. Its remarkable and puzzling multiwavelength (MWL) characteristics raise fundamental questions about the interplay between the jet and accretion disk, as well as the enigmatic nature of (variable) gamma-ray emission in steep-spectrum radio sources with jets not pointing at the observer.

After a summary of the intriguing MWL characteristics of 3C 286, combining the properties of compact steep-spectrum source (CSS), very radio-loud Narrow-line Seyfert 1 (NLS1) galaxy, gamma-ray detection despite a misaligned jet, and presence of a damped Lyman alpha absorber (DLA) despite an apparent lack of X-ray absorption, rarely seen together in a single system, we present new observations across the electromagnetic spectrum from  the radio to high-energy regime and discuss several key open questions: how much variability in each spectral band, how much reddening and X-ray absorption, and which mechanism drives the gamma-ray emission  ? 

We use a cosmology with 
$H_{\rm 0}$=70 km\,s$^{-1}$\,Mpc$^{-1}$, $\Omega_{\rm M}$=0.3 and 
$\Omega_{\Lambda}$=0.7
and the cosmology calculator of \citet{Wright2006}. This corresponds to  
a scale of 7.66 kpc/'' at the redshift of 3C 286 ($z$=0.849). 

\subsection{Radio properties and radio calibrator}

3C 286 is part of the 3rd Cambridge Catalogue of radio sources from 1959 \citep{Edge1959}. After an improved radio position was taken with the Owens Valley Radio Observatory in the early 1960s, 
it was possible to identify its optical counterpart  initially thought to be a star \citep{Matthews1963}. 

Later, 3C 286 was  one of the first quasars imaged with the technique of Very Long Baseline Interferometry (VLBI). 
Radio images at sub-arcsecond resolution revealed a core region  with a jet and counter jet,  
dominated by the central component \citep{Wilkinson1979, Pearson1980, Spencer1989}. 
On large scales, two lobes extend 0.7'' to the East and 2.6'' to the South-West \citep{Spencer1989,vanBreugel1992, An2004}. 
At tens-of-parsec-scale resolution, the central region is resolved into two bright components of comparable flux density (at 15 GHz), interpreted as two hot spots or a core--jet structure \citep{Zhang1994, Cotton1997, Kellermann1998, An2017, Lister2018-mojave}. 
\citet{An2017} favored the core--jet interpretation, where the NE component (C1) with an inverted spectrum below $\sim 8$ GHz represents the core, and the SW component (C2) represents the brightest knot in the jet. Based on that interpretation, our viewing angle towards the inner jet of 3C 286 on tens-of-parsecs scale was estimated at 48$^{\rm o}$. The inner jet is only mildly relativistic with $v \sim 0.5$c \citep{An2017}. 

The (integrated) radio spectrum of 3C 286 is steep between 1.4 and 50 GHz ($\alpha =-0.6$; $S_\nu \propto \nu^{\alpha}$). Given its compactness, 3C 286 was therefore identified as CSS \citep{PeacockWall1982}. 

3C 286 is one of the strongest extragalactic radio sources in polarized emission \citep{RudnickJone1983, Akujor1995} ($\sim$1 Jy at 5 GHz).  
Its fractional polarization, about 10\% at 20 cm, increases with increasing frequency up to 17\% at 1.3 mm observed with ALMA \citep{Nagai2016}.  

Its very constant (integrated) radio properties established 3C 286 as a radio calibrator in both total flux and linear polarization \citep[e.g.,][]{
Ott1994, Perley2013, Perley2013b}. 

\begin{table}[H]
\caption{Properties of 3C 286. Column entries are as follows: (1) Source coordinates in right ascension (RA) and declination (Decl), (2) redshift, (3) SDSS g magnitude, (4) H$\beta$ line  width, (5) SMBH mass, (6) Eddington ratio, and (6) radio loudness at 5 GHz. }
\label{tab:basic-prop}
\scriptsize
\centering
\begin{tabular}{cccccccc}
\toprule
J2000 coordinates & $z$ &  g magnitude & FWHM(H$\beta_{\rm broad}$) & SMBH mass  & $L/L_{\rm Edd}$  & radio loudness\\
	RA, Decl      &    & & \kms\, &  M$_{\odot}$  & & $R_{\rm 5 \,GHz}$   \\
(1) & (2) & (3) & (4) & (5) & (6) & (7) \\
\midrule
13h31m08.2879s, +30$^o$30'32.958''  & 0.849 & 17.33 & 2001 & 1.3\,10$^{8}$ & 1.0 &  10$^{4.4}$ \\
\bottomrule
\end{tabular}
\end{table}

\subsection{Optical spectroscopy and NLS1 classification}

The optical spectrum of 3C 286 exhibits bright permitted and forbidden emission lines at redshift $z=0.849$ \citep[][our Tab. \ref{tab:basic-prop}]{Burbidge1969, Hirst2003, Schneider2007}. 
First optical spectra were taken by Maarten Schmidt in 1962, when 3C 286 was still considered a 'radio star' and its extragalactic nature was not yet known \citep{Schmidt1962}. Three years later the first redshift measurement of 3C 286 was published \cite{Oke1965}, implying its cosmological distance. 

The NLS1 classification{\footnote{Formally, 3C 286 is a narrow-line type 1 quasar. However, we follow the approach common in the literature to refer to all narrow-line type 1 AGN (quasars and their lower luminosity equivalents of Seyfert galaxies) collectively as NLS1 galaxies.}} of 3C 286 was first suggested in 2016, based on new SDSS \citep{Dawson2013} spectroscopy which included the H$\beta$--[OIII] regime for the first time at high signal-to-noise \citep{Yao2016, Berton2017}.  A detailed analysis of the optical spectrum was  presented by \citet{Yao2021}, firmly establishing the NLS1 classification. 
NLS1 galaxies are defined as AGN with narrow widths of their Balmer lines from the broad-line region (BLR) with FWHM(H$\beta_{\rm broad}$)$<$2000 km/s, [OIII]5007/H$\beta_{\rm totl}<3$, and FeII/H$\beta_{\rm totl}>0.5$  \citep{OsterbrockPogge1985, Goodrich1989, Veron2001}. They show a host of exceptional MWL characteristics, at one extreme end of AGN correlation space \citep{Sulentic2000a, Boroson2002, Xu2012, Grupe2010} (for recent reviews see: NLS1 properties in the MWL bands: \citep{Komossa2008-rev}, the X-rays: \citep{Gallo2018}, the radio: \citep{Lister2018}, the gamma-rays: \citep{Dammando2019, Paliya2019}, and the MWL properties of the radio-loud sources: \citep{Komossa2018}).  
Initially mostly studied in the optical \citep{OsterbrockPogge1985} and X-ray \citep{Puchnarewicz1992} regime, the radio properties of NLS1s revealed that only 7\% of them are radio-loud \citep{Komossa2006} with a mix of CSS and blazar-like sources \citep{Komossa2006, Yuan2008}. A subset of them was detected in the gamma-ray regime \citep{Abdo2009-nls1}, some highly variable re-confirming the presence of relativistic jets. 
Perhaps the most remarkable recent finding was the detection of high-amplitude radio flaring of several previously radio-silent NLS1 galaxies \citep{Anne2018, Berton2020, Emilia2023}.

3C 286 fulfills all of the optical spectroscopic NLS1 galaxy classification criteria: an FWHM(H$\beta_{\rm broad}$) = 2000 km/s{\footnote{Note that strong projection effects are unlikely to affect the Balmer-line profile, since 3C 286 is misaligned \citep{Berton2017}, i.e. it is not a blazar viewed near face-on. We further note that 3C 286, like many NLS1 galaxies \citep{Sulentic2000b, Veron2001, Xu2012} has broad wings in its Balmer lines, and its H$\beta_{\rm broad}$ profile is alternatively well fit by a Lorentz profile (rather than double Gau{\ss} profiles) with FWHM(H$\beta_{\rm broad, L}$)=1858 km/s.}}, a line ratio [OIII]5007/H$\beta_{\rm totl}=1.1$, and a line ratio FeII\,4570/H$\beta_{\rm totl}$=0.76, where FeII\,4570 is calculated by integrating the FeII complex between 4434 and 4684 \AA.
In the quasar main-sequence diagram \citep[FWHM(H$\beta$) vs FeII/H$\beta$;][]{Sulentic2000b, Marziani2018}, 3C 286 is located well within the region of  population A sources (category A2). 

The high-ionization lines in the optical spectrum of 3C 286, [OIII]5007, [NeIII]3870, and [NeV]3426, exhibit strong blueshifted components (blue wings) which are highly broadened up to 2398 km/s FWHM in case of [NeIII] \citep{Yao2021}, indicating strong outflows and/or jet-cloud interactions. 
Based on its FWHM(H$\beta_{\rm broad}$) 3C 286 hosts an SMBH with mass 1.3 $\times 10^{8}$ M$_\odot$, and accretes at an Eddington ratio $L/L_{\rm Edd}=1$ \citep{Yao2021} consistent with the value of  $L/L_{\rm Edd}\sim0.9$ obtained from SED modelling \citep{Zhang2020}.

The spectral energy distribution (SED) of 3C 286 rises steeply in the optical \citep{Zhang2020, Yao2021}, interpreted as emission from the accretion disk dominating over the non-thermal emission from the jet at optical-UV wavelengths.  

3C 286 is very radio-loud \cite{Yao2021}. Given its  Kellermann \citep{Kellermann1989} radio-loudness index of $R_{\rm{5\,GHz}}$ = $f_{\rm  5\,GHz}/f_{\rm{4400 A}}$ = 10$^{4.4}$, 3C 286 is among the radio-loudest NLS1 galaxies known to date  (see for comparison, Fig. 8 of the largest SDSS NLS1 sample;  \citep{Paliya2024}).

\subsection{UV properties and DLA system}
The UV spectrum of 3C 286 shows a damped Lyman alpha absorber (DLA) system at redshift $z=0.692$ \citep{Cohen1994}. A foreground low-surface brightness galaxy at $\sim$2.5'' distance from 3C 286 \citep{Steidel1994, LeBrun1998} is thought to be responsible for the UV absorption as well as for the HI 21 cm absorption detected at that redshift \citep{BrownRoberts1973}.  Based on the UV spectrum, a column density of the intervening absorber of $N_{\rm H} \sim 2 \times 10^{21}$ cm$^{-2}$ was estimated \citep{Cohen1994}. 

\subsection{Fermi gamma-ray detection of 3C 286 despite its misaligned jet}

3C 286 is associated with the Fermi-LAT gamma-ray source 4FGL\,J1331.0+3032 \citep{Ackermann2015, Ballet2024}. 
Based on the first 11 yrs of Fermi observations, \citet{Zhang2020} 
reported 
an average gamma-ray flux in the energy range 100 MeV--300 GeV of
$f_{\gamma}=2.32 \times 10^{-12}$ erg\,cm$^{-2}$\,s$^{-1}$. The  significance of the gamma-ray signal, based on the maximum  likelihood Test Statistics (TS) is TS=54.2, and there is evidence for gamma-ray variability at TS$_{\rm  var}$=64.3, where  TS$_{\rm var}$ is the variability index \citep{Zhang2020}. 
The evidence for variability depends on the bin size of the light curve, and lower values of  TS$_{\rm  var}$, consistent with constant   emission, have  also been reported in the literature \citep{Principe2021, Ballet2024}.      

The majority of long-lived extragalactic high-energy emitters detected in the gamma-ray regime are blazars \citep{Fichtel1994, Ackermann2015, Foschini2022}. 
3C 286 is one of only a small fraction of misaligned sources 
detected with Fermi, raising the question about the gamma-ray emission mechanism {\footnote{At an association probability of 0.985 \citep{Ballet2024}, 3C 286 is the most likely counterpart of the Fermi source, but there remains a small possibility of an alternative             counterpart identification.}.  A variety of different models have been discussed in the literature
but no single  emission mechanism valid   for the whole population has been robustly established yet (see our  Sect. 6.5).  

\subsection{X-ray properties}

The gamma-ray detected NLS1 galaxies show a range of spectral shapes in X-rays \citep{Foschini2015, Dammando2020, Yao2023}. Some are dominated by a flat jet component, while others show a significant soft excess commonly interpreted as emission from the accretion disk. 

Until recently, 3C 286 was rarely observed in the X-ray band. It was detected with Einstein \citep{Tananbaum1983} and in the ROSAT all-sky survey \citep{Voges2000}. 1--2 ks snapshots with Chandra \citep{Massaro2015, Yao2021} and the Neil Gehrels Swift observatory \citep{Yao2021} (Swift hereafter) revealed a flat X-ray spectrum with $\Gamma_{\rm x}$ $\approx 2$, but more complex spectral models were not constrained and could not be excluded. \citet{Yao2021} also reported evidence for X-ray  variability, raising the question if the X-rays were dominated by the disk or jet; important in particular because of  the use of 3C 286 as radio calibrator.   

Given its remarkable MWL properties,  and the evidence for X-ray variability in particular, we obtained new observations with Swift, the Effelsberg radio telescope, and Chandra that we present in the next sections. Tab. \ref{tab:class} provides a summary of the main classifications of 3C 286.

\begin{table}[H]
\caption{MWL classifications of 3C 286.}\label{tab:class}
\begin{tabular}{lll}
\toprule
classification/property & waveband & comment \\
\midrule
NLS1 galaxy & optical &  among radio-loudest known  \\
CSS         & radio   &  spectral index $\alpha =-0.6$  \\ 
misaligned Fermi source & gamma-rays &   \\ 
DLA absorber  & UV/radio &  intervening; $z=0.692$ \\ 
\bottomrule
\end{tabular}
\end{table}

\section{Swift observations}
\label{sec:swift}

In order to search for X-ray flux and spectral variability of 3C 286 on both short and long time scales of days to years, we have observed the quasar seven times with Swift \citep{Gehrels2004}  between 2020 August and 2024 April (Tab. \ref{tab:obs-log}). 

\begin{table}[H]
\caption{Log of our Swift observations of 3C 286 between 2020 and 2023  with observation ids (OBSIDs) 13644--1 to 13644--8 (except 13644--5, not carried out). The Swift instrument is listed in the first column, the UVOT filters are reported in the second column, the energy band of the XRT and  the central wavelengths of the UVOT filters \citep{Poole2008} are given in the third column, and the exposure times $t$ are reported in the remaining columns. 	
}
	\label{tab:obs-log}
 \scriptsize
	\begin{tabular}{lclccccccc}
		\toprule
		 & filter & $\Delta E$ or $\lambda_c$ & $t$ (ks) & &&&& & \\
             &  &  & 2020-08-16 & 2020-08-21 & 2023-11-04 & 2023-11-14 & 2023-11-29 & 2023-12-03 & 2024-04-18\\ 
		\midrule
		XRT &  & 0.3--10 keV & 2.03 & 1.39 & 1.93 & 1.68 & 1.66 & 2.18 & 1.51\\
        UVOT & W2 & 1928 \AA  & 0.65 & 0.42 & 0.63 & 0.55 & 0.57 & 0.71 & 0.48\\
             & M2 & 2246 \AA & 0.49 & 0.37 & 0.45 & 0.38 & 0.23 & 0.12 &0.33 \\
             & W1 & 2600 \AA  & 0.32 & 0.21 & 0.31 & 0.28 & 0.35 & 0.35 &0.24\\
             & U & 3465 \AA & 0.16 & 0.10 & 0.16 & 0.14 & 0.10 & 0.18 &0.12\\
             & B & 4392 \AA & 0.16 & 0.10 & 0.16 & 0.14 & 0.10 & 0.18 &0.12\\
             & V & 5468 \AA & 0.16 & 0.10 & 0.16 & 0.14 & 0.10 & 0.18 & 0.12\\
		\bottomrule
	\end{tabular}
\end{table}

\subsection{XRT data reduction and spectral analysis}

The  X-ray data were obtained with the Swift X-ray telescope \citep[XRT;][]{Burrows2005} operating in photon counting mode \citep{Hill2004}. To carry out the timing and spectral analysis, we selected source photons within a circular area with a radius of 20 detector pixels  (one pixel corresponds to a scale of 2.36''). Background photons were extracted in a nearby circular region with an extraction radius of 100 detector pixels. X-ray spectra of the source and background in the energy band 0.3--10 keV were generated. The spectral analysis was then carried out with the package XSPEC, version 12.14.0b \citep{Arnaud1996}. 
The  X-ray light curve is shown in Fig. \ref{fig:lc_swift}. 
Spectral fits were performed on the unbinned data and based on W-statistics \citep{Cash1979, Arnaud2024} (W-stat hereafter).

  \begin{figure}[t]
   \centering
  \includegraphics[clip, trim=0.9cm 5.3cm 1.0cm 1.3cm, width=12.5cm]{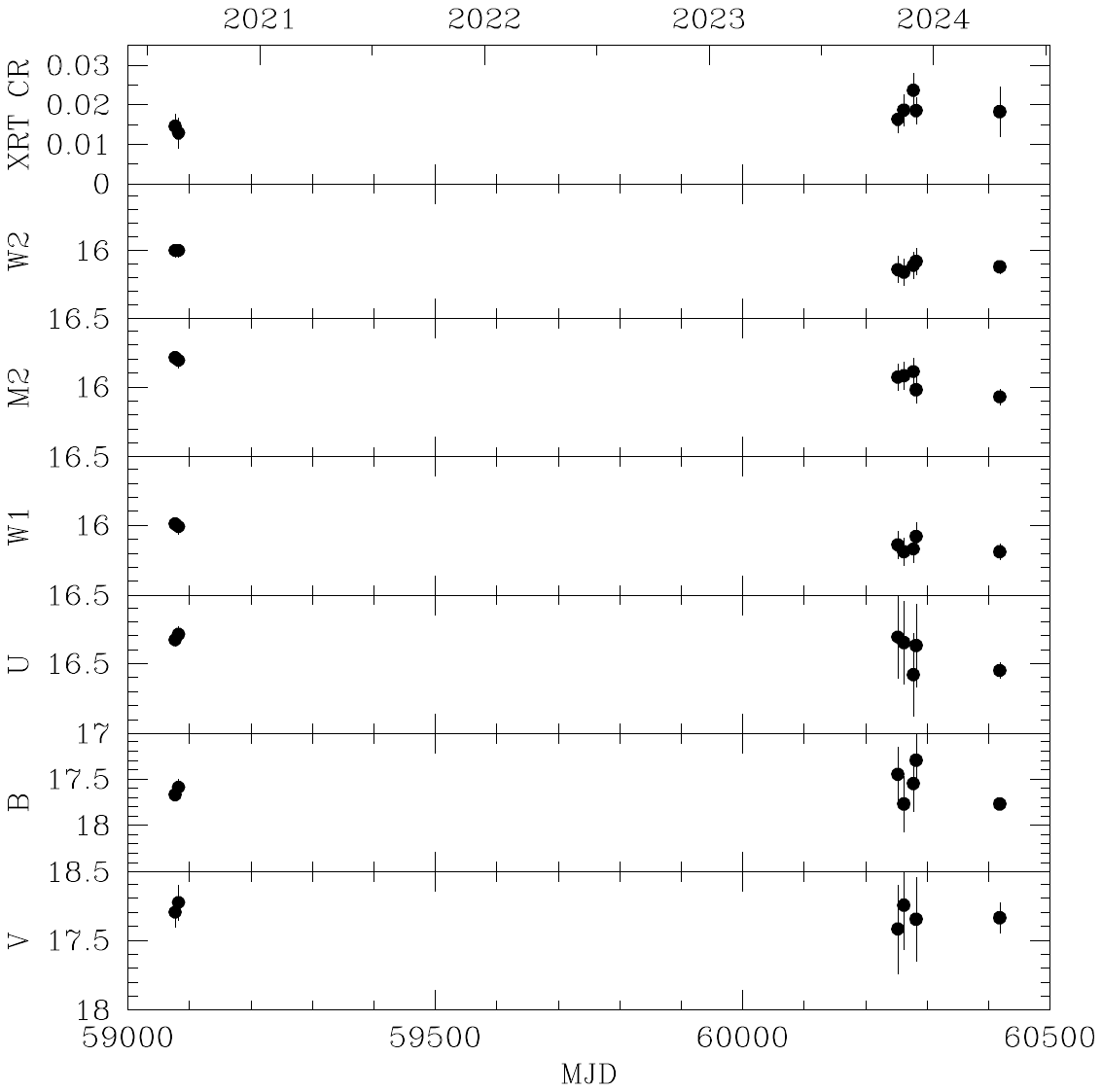}
   \caption{Swift light curve of 3C 286 between 2020 and 2024. UV--optical magnitudes were corrected for {\bf{Galactic}} extinction. 
}
\label{fig:lc_swift}%
    \end{figure}

In order to measure the average X-ray spectral slopes of 3C 286, we have split the Swift observations of 3C 286 into low-state (2020) and high-state (2023) data and then merged all data sets of each state in order to obtain two spectra. For the auxiliary response files (arfs) we applied the FTOOLS command {\it addarf} and weight the contribution of each arf by the exposure time of the individual spectra divided by the total exposure times of the merged spectra. The response file {\it swxpc0to12s6--20130101v014.rmf} was applied to the spectra. 

Both spectra were then first fit with a single powerlaw subject to Galactic absorption of $N_{\rm H} = 1.11 \times 10^{20}$ cm$^{-2}$ \citep{HI4PICollaboration2016}. This gives a powerlaw photon index of $\Gamma_{\rm x}$ = 1.83$\pm{0.27}$ (W-stat/d.o.f. = 32.9/36) in low-state and $\Gamma_{\rm x}$ = 2.11$\pm{0.15}$ (W-stat/d.o.f. = 72.4/90) in high-state (Fig. \ref{fig:spec_swift}).  If the column density $N_{\rm H}$  is left as free parameter at $z=0$, it underpredicts the Galactic value but comes with large uncertainies and was therefore fixed at the Galactic  $N_{\rm H}$.
Given the differences in the two spectra, next, both were fit simultaneously. The spectral index $\Gamma_{\rm x}$  was tied to the same value in this approach, and the normalizations were left free to vary. This again implies significant variability, with $K_{\rm pl} = 2.6\pm{0.5} \times 10^{-4}$ (2020) and $4.1\pm{0.5} \times 10^{-4}$ (2023), where $K_{\rm pl}$ is the normalization at 1 keV in units of photons\,keV$^{-1}$\,cm$^{-2}$\,s$^{-1}$. 

Next, in order to test for more realistic and complex spectral models and evaluate the cause for variability, 
we have used a powerlaw plus soft excess description. 
The soft excess was parameterized as blackbody, and additional cold absorption from an intervening absorber at  redshift $z=0.692$ and with $N_{\rm H} = 1.75 \times 10^{21}$ cm$^{-2}$ was added. 
The XSPEC model used was {\it TBabs * zTBabs * (zpowerlaw + zbbdy)}, where {\it TBabs} and {\it zTBabs} are the absorption models as described by \citet{Wilms2000} in our Galaxy and at a redshift $z=0.692$, respectively, {\it zpowerlaw} is the powerlaw model at $z=0.849$, and {\it zbbdy} is the blackbody model at $z=0.849$. The powerlaw model is defined as 
$A(E)=K_{\rm pl} [E(1+z)]^{-\Gamma_{\rm x}}$, where $K_{\rm pl}$ is the normalization at 1 keV in units of photons\,keV$^{-1}$\,cm$^{-2}$\,s$^{-1}$. The blackbody model is defined as 
$A(E)$ = $K_{\rm bb}$ ${8.0525[E(1+z)]^2{\rm d}E}\over{(1+z)(kT)^4[\exp[E(1+z)/kT]-1]}$, 
where $kT$ is the temperature in keV, and $K_{\rm bb}$ is the normalization in 
$L_{39}/[D_{10}(1+z)]^{2}$, where $L_{39}$ is the source luminosity in units of 10$^{39}$ erg\,s$^{-1}$ and $D_{10}$ is the distance to the source in units of 10 kpc \citep{Arnaud2024}.
Since multi-parameter fitting is not warranted for the short Swift observations, we have fixed the intervening absorption, powerlaw index, and the temperature of the blackbody component to the preliminary values obtained from our recent XMM-Newton observation of 3C 286 \citep{Yao2024}: $\Gamma_{\rm x}$=2.09 and  $kT_{\rm bb}$=0.177 keV, and left the normalizations of both components free to vary. The difference between the two spectra can then be tentatively explained by an increase in the soft (blackbody) component in 2023. Results of the spectral fits are reported in Tab. \ref{tab:spectral-fits}. 
The XMM-Newton spectral fit results based on the same model are shown for comparison.
With Swift, the following 0.3--10 kev fluxes (corrected for Galactic absorption and for intervening absorption) of the powerlaw plus blackbody model are measured: $f$=5.1 $\times 10^{-13}$ erg cm$^{-2}$ s$^{-1}$ (2020) and 8.3 $\times 10^{-13}$ erg cm$^{-2}$ s$^{-1}$ (2023), respectively. 
[For comparison, the simple single powerlaw fit gives $f$=4.5 $\times 10^{-13}$ erg cm$^{-2}$ s$^{-1}$ (2020) and 6.3 $\times 10^{-13}$ erg cm$^{-2}$ s$^{-1}$ (2023), respectively. These values are lower than from the two-component fit, because they miss the blackbody component and the correction for the extra absorption.]

   \begin{figure}[t]
   \centering
   \includegraphics[width=10cm]{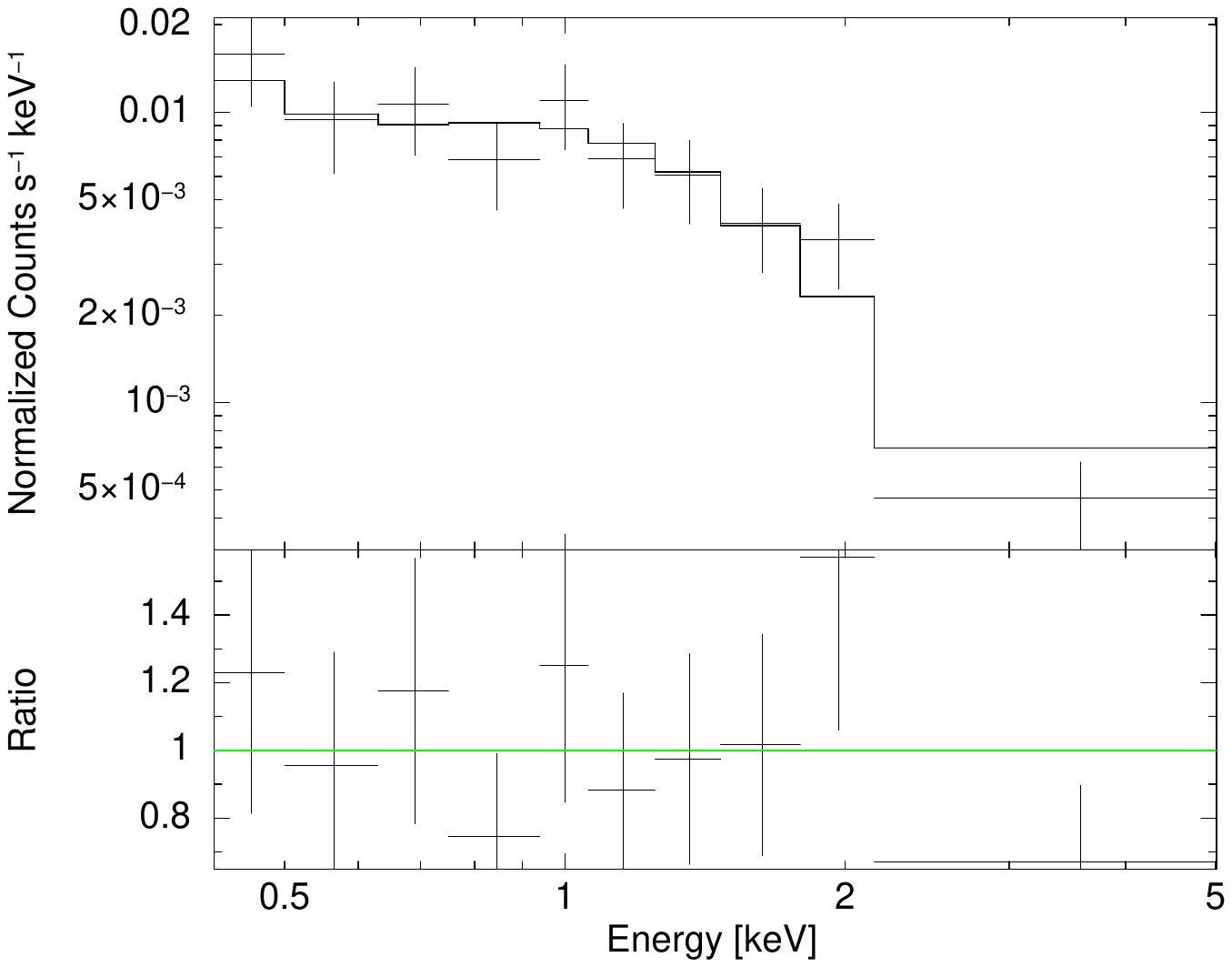}
   \caption{Composite binned Swift X-ray spectrum of 3C 286 from 2023 based on a single powerlaw fit. (Note that the actual spectral fits were done on the unbinned spectrum.) 
   }
              \label{fig:spec_swift}%
    \end{figure}

A source detection was performed within XIMAGE
on the co-added 2020--2023 data sets. 
All other X-ray sources are much fainter than 3C 286 itself. 
In Sect. 8.2 of the Appendix, all sources detected above 3$\sigma$ are listed. 
                     
\begin{table}
\centering
\caption{Results of Swift X-ray spectral fits to 3C 286 for the 2020 low-state and 2023 high-state. XMM-Newton results are shown for comparison. Columns are as follows: (1): column density of the Galactic absorption, (2) column density of the intervening absorber at $z$=0.692, (3) X-ray powerlaw photon index $\Gamma_{\rm x}$, (4) powerlaw  normalization $K_{\rm pl}$ at 1 keV  in units of photons\,keV$^{-1}$\,cm$^{-2}$\,s$^{-1}$, (6) blackbody temperature $kT$ in eV, and (6) normalization $K_{\rm bb}$ of the blackbody component in units of  
$L_{39}/[D_{10}(1+z)]^{2}$ (see Sect. 2.1). } 	
 \label{tab:spectral-fits}
	\begin{tabular}{lllllll}
		\toprule
	date	& N$_{\rm{Gal}}$ & N$_{z=0.692}$ & $\Gamma_{\rm x}$ & $K_{\rm pl}$ & $kT$ & $K_{\rm bb}$ \\
       & 10$^{20}$ cm$^{-2}$ & 10$^{21}$ cm$^{-2}$ & & 10$^{-4}$ & eV & 10$^{-5}$ \\ 
        & (1) & (2) & (3) & (4) & (5) & (6) \\ 
		\midrule
   2020 low-state & 1.11 & -- & 1.83$\pm{0.27}$ & 2.23$\pm{0.63}$& -- & -- \\
                 & 1.11 & 1.75 & 2.09 & 3.03$\pm{0.67}$ & 177 & 0.42$^{+0.59}_{-0.42}$ \\
   2023 high-state & 1.11 & -- & 2.11$\pm{0.15}$ & 4.31$\pm{0.64}$ & -- & -- \\
                 & 1.11 & 1.75 & 2.09 & 4.21$\pm{0.55}$ & 177& 1.43$\pm{0.60}$ \\
                 \\
    2023 XMM & 1.11 & 1.75$^{+0.62}_{-0.49}$ & 2.09$\pm{0.10}$ & 3.40$_{-0.46}^{+0.49}$& 177$^{+28}_{-23}$ & 1.13$_{-0.43}^{+0.92}$ \\
    \bottomrule
	\end{tabular}
\end{table}

\subsection{UVOT data reduction and analysis}

The UV-optical telescope \citep[UVOT;][]{Roming2005} was used to observe 3C 286 seven times in three optical and three UV filters.
For further analysis, the data sets of each filter were first co-added in each single observation. The source counts were then extracted in a region of circular or elliptical shape centered on 3C 286.  A nearby area of 20'' radius was used to extract the  
background counts. 
The radius of the source extraction region was the standard 5'' during the 2020  observations. However, the data in 2023   are affected by a slight, uncorrected drift motion of the Swift satellite \citep{Cenko2024} such that sources appear no longer point-like but slightly elongated.

Therefore, different source extraction methods and tests were performed.  
First, a larger source extraction region of 15'' radius was used. Overall, this gives values consistent with the 2020 measurements, except for occasional single-filter deviations by up to 0.3 mag which are not physical. 
Next, we have used elliptical source extraction regions of varying radius and orientation,  individually matching the source elongation in each filter and observation. This gives very similar results. Next, we have checked if any of the UVOT observations fell onto the known small-scale low-sensitivity areas (\url{https://www.swift.ac.uk/analysis/uvot/sss.php}) of the detector, but that was not the case.  Finally, we have also checked photometry of several other brighter and fainter sources in the field of view, and overall their emission is constant. 
In conclusion: While high-precision photometry is not possible for the 2023 epochs, the optical and UV magnitudes of 3C 286 are found to be consistent with being constant within an uncertainty of $\pm$0.3 mag in the optical, and $\pm$0.1 mag in the UV bands. 

The  background-corrected counts were converted into magnitudes in the Vega system  using the latest calibration as described by \citet{Poole2008} and \citet{Breeveld2010}. The magnitudes 
in each filter were
determined with the UVOT tool {\it uvotsource} with the parameter {\it apercorr} set to {\it curveofgrowth} in order to compensate for the deviations from the calibrated circular 5'' radius source extraction region in 2023.
 The data were corrected for Galactic reddening
based on the reddening curves of \citet{Cardelli1989} and using $E_{\rm B-V}$=0.01. 
The extinction-corrected magnitudes are reported in Sect. 8.1 of the Appendix.

\section{Effelsberg radio flux density and polarimetry observations}
\label{sec:radio}

3C 286 was observed on 2023 November 3 and 16 in the radio regime, quasi-simultaneous with the Swift X-ray, UV and optical observations of November 4 and 14.
Observations of 3C 286 with the Effelsberg 100m radio telescope 
are routinely performed within the MOMO project \citep[Multi-wavelength Observations and Modelling of OJ 287; e.g.][] {Komossa2021, Komossa2023}. 
Usually, 3C 286 serves as one of several radio calibrators during these observations, and its constant radio emission is always carefully checked
against other calibration observations and against  the theoretically  expected telescope sensitivity. 
However, for the analysis of the data in 2023 November, 3C 286 was treated as the main target, and 
additional other radio calibrators (e.g., 3C161, NGC 7027) were observed and used to measure the flux density and polarisation of 3C 286 itself, in order to re-confirm 
the constancy of the radio flux of 3C 286  
quasi-simultaneous with the new Swift MWL data. 

\subsection{Flux densities}
Flux densities were measured between 2.6 and 41 GHz. Data acquisition and reduction
was carried out as follows \citep{Kraus2003}.
The observations were carried out in form of cross-scans.  Several sub-scans were performed in azimuth and elevation over the source position. The amplitude (antenna temperature) and pointing offsets were determined by fitting a Gaussian profile to the individual sub-scans. After correction for the usually small pointing offset, the amplitudes of all sub-scans were averaged.
In the next step, the atmospheric attenuation as well as the telescope's gain-elevation effect (loss of sensitivity due to the gravitational deformation of the main dish when tilted) were corrected. Finally, the corrected antenna temperatures were compared to the expected flux densities using the calibrator sources and telescope sensitivities for the conversion into the Jansky scale. 
Typical errors are in the range of 0.5-2\% for the lower frequencies
and about 5-10\% for the high frequencies (depending mostly on the weather conditions).

\subsection{Polarimetry}
Polarimetry was conducted with the Effelsberg telescope at several frequencies between 2.6 and 41 GHz.
For the analysis of the linear polarization the values of  Stokes $Q$ and $U$
were taken from the corresponding channels of the cross-scan data at the position
of the maximum of the Stokes $I$ channel.
For the correction of the instrumental effects, the Müller matrix-method
\citep[e.g.,][]{Turlo1985} has been applied using average Müller matrices from the calibration data base; i.e. from former observations. As with Stokes I, additional calibrators have been used to check the consistency of the results.
Typical measurement errors are $\sim$5\% for the polarized flux density $P$ and between  3$^{\rm o}$--5$^{\rm o}$ for the polarization angle $\chi$  (depending on the frequency and weather conditions).

Results (Fig. \ref{fig:radio}) show that the radio emission of 3C 286 
is constant, when comparing with 
past radio measurements of 3C 286 at similar  resolution. No variability is detected in the radio regime. The fractional polarization $p$ 
shows a small dependence on frequency increasing from 11 to 13\% (see Sect. 8.1 of the Appendix). 

   \begin{figure}[]
  \centering
   \includegraphics[width=15.0cm]{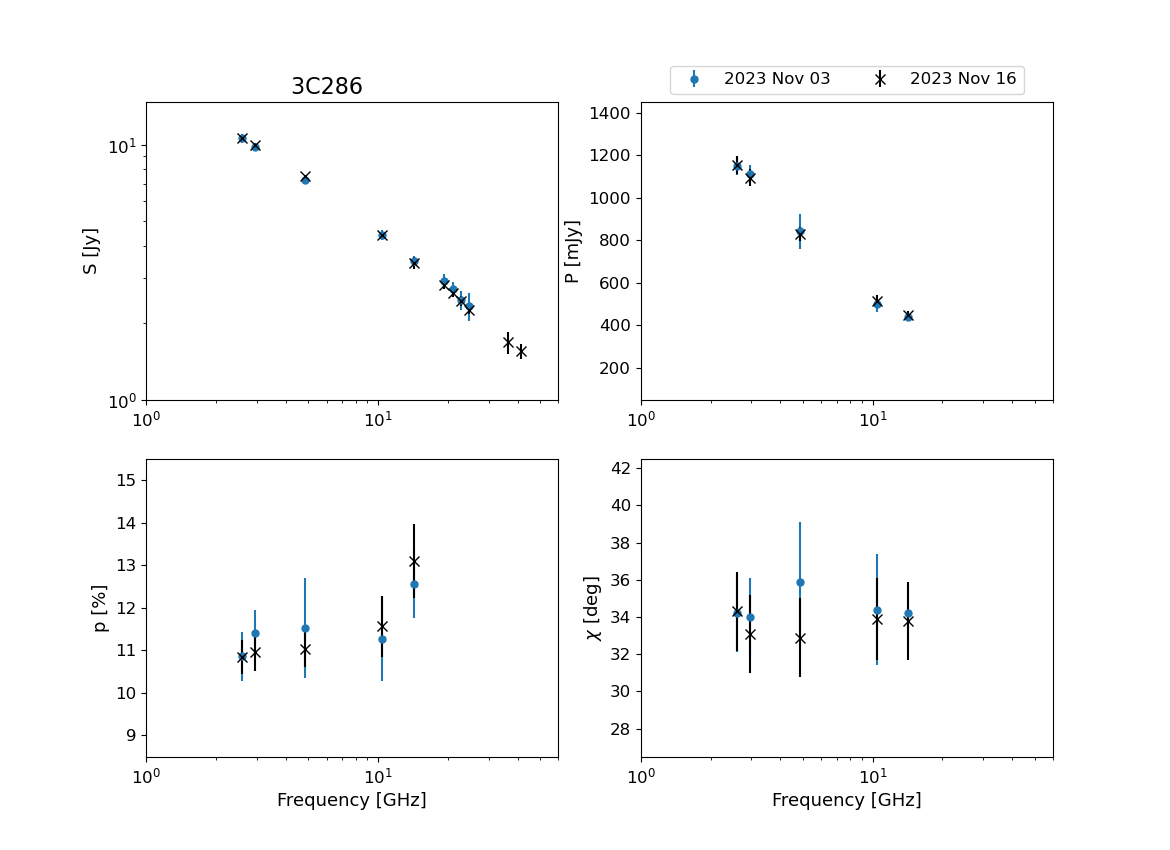}
   \caption{Radio observations of 3C 286 from 2023 November. $S$ is the flux density, $P$ the polarized flux density, 
   $p$ the polarization percentage, and $\chi$ the polarization position angle. Measurements from 2023 November 3 are marked in blue, measurements from November 16 in black. 
   }
              \label{fig:radio}%
    \end{figure}

\section{Chandra imaging observation}

An observation of 3C 286 with the Chandra ACIS-S detector \citep{Weisskopf2002} was performed on 2023 February 12 with a duration of 35 ks. The data will be discussed in more detail by \citet{Yao2024}, but we use them here to search for additional sources in the vicinity of 3C 286 separately in the soft and hard X-ray band (Sect. 6). 

\section{Longterm X-ray light curve}

To construct the longterm X-ray light curve of 3C 286, data from previous X-ray missions were collected and converted into fluxes and then into Swift XRT count rates in the band 0.3--10 keV assuming a constant spectral shape consisting of the single powerlaw description with Galactic absorption and $\Gamma_{\rm X}$=2.1. Count rate data are from Einstein \citep{Tananbaum1983}, ROSAT \citep{Voges2000}, Chandra \citep{Massaro2015, Yao2024}, and XMM-Newton \citep{Yao2024}, in addition to the Swift data itself. We have always used the directly measured count rates for conversion into fluxes as the previously published fluxes depend on the assumption of the spectral model and the amount  of absorption. The single exception are the Chandra   observations which are subject to a decrease in the  ACIS low energy quantum efficiency caused by increasing contamination on the Optical Blocking Filter, implying a decrease of  the directly  measured count rate. Instead, we used the measured powerlaw flux for conversion. Data from the very short 2 ks 2013 observation are from \citet{Massaro2015}. 
For the Swift data  of 2020 and 2023, the average count rate is shown. 
 3C 286 shows constant baseline emission between  
 1990 (ROSAT) and the 2020s (XMM-Newton), and evidence for flaring up to a factor of $\sim$2 with Swift (Fig. \ref{fig:longlight}). 

   \begin{figure}[H]
   \centering
   \includegraphics[width=10cm]{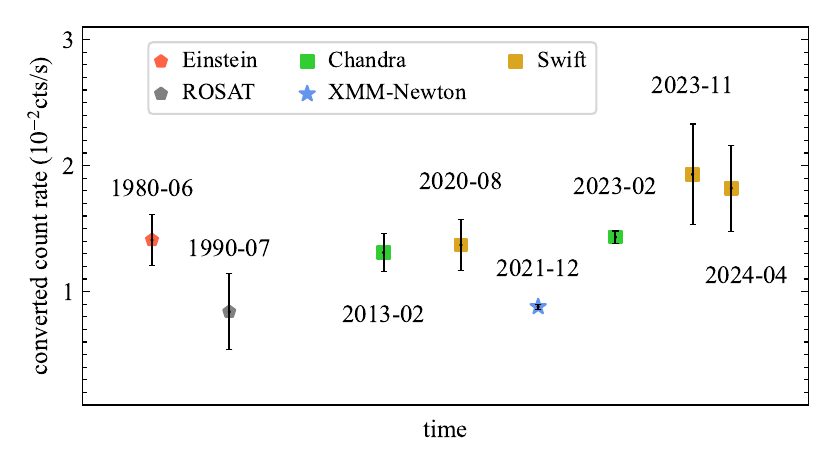}
   \caption{Longterm X-ray light curve of 3C 286 (displayed as Swift XRT count rate in the band 0.3--10 keV), observed with multiple missions between 1980 and 2024 (note that the time axis is not linear because of the widely spaced observations). All non-Swift data were first converted into fluxes in a homogeneous way and then into Swift XRT count rates  (see text for details). 
   }
              \label{fig:longlight}%
    \end{figure}

\section{Discussion}

We have presented  new Swift optical, UV and X-ray observations, new  radio data taken at the Effelsberg 100m radio telescope, and the first deep  imaging  observations of 3C 286 in the X-ray regime (the latter will be   further discussed by \citep{Yao2024}).  These data are used to address some of the open questions regarding the intriguing MWL properties of 3C 286. 

\subsection{How much radio variability ?}

3C 286 is a well-known radio calibrator and the constancy of its flux density and polarization in the past has been well established \citep{Ott1994, Peng2000, Agudo2012, Perley2013, Perley2013b, Perley2017}. At the same time,  recent evidence for X-ray variability \citep{Yao2021} and evidence for radio core variability \citep{An2017} has raised the question, if 3C 286  is still constant in the radio regime (in low-resolution observations), and especially, if the X-ray variability is driven by the disk or the jet.
We have re-observed 3C 286 with the Effelsberg telescope in 2023 November quasi-simultaneous with Swift, in order to search for any possible radio variability. 

Results confirm the constancy of the radio properties of 3C 286 at the Effelsberg telescope resolution. The measured flux densities were compared with past radio observations of 3C 286 at comparable resolution \citep[e.g.,][]{Kellermann1969, Kuehr1981, Ott1994, Gabanyi2007, Massardi2009, Mantovani2009, Richards2011} and values agree within the error bars. 
Further, the new observations show the well-known (small) increase in fractional polarization $p$ 
with frequency (e.g., \citep{Perley2013, Nagai2016, Hull2016}; see also Fig. 3 of \citep{Pasetto2018}). We measured a change of $p$ from 11 to 13\% as frequency increases. The polarization angle
 $\chi$ of 33--34$^{\rm o}$ 
 is consistent with previous measurements as well. 

\subsection{X-ray spectrum, soft excess, and variability}

The X-ray spectral results are of interest for several reasons: disentangling the disk/jet contributions, constraining the amount of absorption along the line-of-sight, and addressing the question, which component it is that varies: jet, disk, or absorber. The previous short X-ray observations of 3C 286 were well fit with a single powerlaw of photon index $\Gamma_{\rm x} \approx 2$ \citep{Yao2021}. Excess absorption as seen  in form of the DLA system was not required, which was a puzzle. However, the new X-ray models have shown, that such  excess absorption from    {\em{intervening}} gas can  be well accommodated when allowing for the presence of a soft excess component.
With the new repeat Swift observations we can now address which component it was that varied, and find that the data are consistent  with a change in the strength of the soft excess which represents the disk component.  It is         interpreted  as  the  tail of the `blue bump' seen in the optical-UV. Such a component is frequently observed in NLS1 galaxies \citep[e.g.,][]{Grupe2010}.

NLS1 galaxies, the majority of them radio-quiet, are known to be highly variable in X-rays \citep{Grupe2010}. It then depends on the variability mechanism(s), whether and how the jet formation and emission at the launching region is affected. For instance, if the X-ray variability was due to partial-covering effects of an ionized disk wind, or arose from hot spots in the disk, or
was due to effects of relativistic reflection, the jet would or could remain unaffected. A number of follow-up observations are essential to constrain further the disk-jet links. For instance, deeper X-ray spectroscopy of 3C 286 will improve measurements of the soft excess, and long-term X-ray monitoring will quantify the amplitude and frequency of variability.   

\subsection{Extinction, absorption, and the very high radio loudness of 3C 286}

Knowledge of the amount of extinction by dust  along our line-of-sight is important for a variety  of reasons: for instance, for deriving and modelling the intrinsic SED, for the optical spectroscopic identification of AGN class, for measuring SMBH mass and Eddington ratio, for constraining the mechanism of X-ray variability, and for determining the radio loudness
(3C 286 is the radio-loudest gamma-ray emitting NLS1 galaxy known. The question is raised, if extinction within 3C 286 itself or along our line-of-sight strongly dims the optical emission, such that, if uncorrected, the radio-loudness index $R$ is highly boosted.) 
There are several ways to evaluate the amount of intrinsic or intervening extinction along our line-of-sight employing 
(1) the BLR Balmer decrement, (2) the NLR Balmer decrement, (3) UV absorption lines (only if the absorber is dusty), (4) 21 cm absorption (only if dusty), and (5) X-ray absorption (only if dusty). 
We comment on each in turn (Tab. \ref{tab:exti}). 

\begin{table}[H]
\centering
\caption{Summary of different measurements of extinction and absorption towards  3C 286.}\label{tab:exti}
\begin{tabular}{lllll}
\toprule
waveband & method & N$_{\rm H}$ &  comment \\
      &    & 10$^{20}$ cm$^{-2}$ &   \\
\midrule
radio & HI 21cm absorption ($z$=0.849) & $<$ 0.052   & \cite{Grasha2019}  \\
radio  & HI 21cm absorption ($z$=0.692)  & &  \citep{BrownRoberts1973} \\ 
optical & BLR Balmer decrement  & ~~ -- & this paper \\ 
optical & NLR Balmer decrement  &   & not yet available \\ 
UV  & Ly $\alpha$ absorption ($z$=0.692) &  ~~~20  &   low dust content \citep{MeyerYork1992, Cohen1994}\\ 
X-rays  &  neutral absorption ($z$=0.692, fixed)& $\sim$17.5 &  \cite{Yao2024}, this paper \\ 
\bottomrule
\end{tabular}
\end{table}

(1) In the optical band, the most reliable of these is the BLR Balmer decrement, because dust is not expected to survive in the bulk of the BLR 
and therefore the extinction must occur along the line of sight, and will affect BLR and continuum emission in the same way. The Balmer decrement of the bluest AGN is close to the case B recombination value \cite{Gaskell2017}. 
Missing H$\alpha$, 
we use the observed H$\gamma$/H$\beta$ flux ratio \citep{Yao2021} of 
$\sim0.36$
which compares to the case B recombination value of 0.46  \citep{Osterbrock1989} and implies little or no extinction within the uncertainties that come with H$\gamma$ line measurement and decomposition. 
(2) The NLR Balmer decrement provides a measure of extinction as well, even though the method does not immediately imply the fraction of dust along our line-of-sight (affecting the continuum) since the dust can be intrinsic to the NLR clouds, not necessarily covering the continuum source.  Since the NLR component of  H$\gamma$ was not yet detected in the optical spectrum, this method has to await future higher S/N spectroscopy. 
(3) UV absorption lines provide a measure of the absorbing gas along the line-of-sight toward the UV continuum source. The absorber will only redden the continuum, if it is dusty. \citet{Cohen1994} estimated a column density of the intervening  Lyman $\alpha$ absorber of 3C 286 of $2 \times 10^{{21}}$ cm$^{-2}$. However, its actual dust content was found to be low, about 5\% of that of the Galactic-disk dust-to-gas ratio \citep{MeyerYork1992}. Therefore, little dust {\em extinction} is expected from this component. 
(4) The 21cm absorption at the redshift of 3C 286 itself was found to be very low, $< 0.052 \times 10^{20}$ cm$^{-2}$ \citep{Grasha2019}, providing a tight constraint on any {\em{intrinsic}} absorption directly along the line-of-sight to the radio continuum source.  
The DLA system at $z=0.692$ is accompanied by HI 21cm absorption at the same redshift. 
As mentioned above, it was argued that the dust content of this component is low. 
(5) X-ray spectroscopy is a potentially powerful way of measuring the amount of absorption along our line-of-sight, even though the dust content itself is usually not constrained (see high-column density dusty {\em ionized} absorbers for an exception \citep{Komossa1998}). Unfortunately, 3C 286 is weak in X-rays due to its high redshift, only medium-deep X-ray observations were obtained so far, and the soft excess and absorption are partially degenerate (Sect. 6.2). The best constraints so far come from the XMM-Newton spectrum of 3C 286 with absorption with a column density comparable to the DLA system \citep[][our Tab. \ref{tab:spectral-fits}]{Yao2024}; intervening not intrinsic. 

In summary, the absorption/extinction intrinsic to 3C 286 is very low, and the intervening absorber is known to have low dust content. 
This finding validates the method used to derive the high Eddington ratio. This finding also reconfirms that the exceptional radio loudness of 3C 286 is not an artefact of heavy extinction affecting the flux ratio 
$f_{\rm  5\,GHz}/f_{\rm{4400 A}}$   used to calculate the radio index $R_{\rm{5\,GHz}}$.
The  very high radio loudness is even more remarkable given that the jet of 3C 286 is not beamed and given that 3C 286 is a CSS so the extended radio emission is rather compact.  

\subsection{Possible detection of X-rays from the SW lobe with Chandra}

The SW lobe is spatially resolved from the core with Chandra, and 3 hard X-ray photons from its region were detected. Given the sharp PSF and low background of Chandra this likely constitutes a detection but deeper imaging is needed to confirm it. 
The lobe X-ray emission may either arise   from upscattered  CMB photons (detected from other radio lobes as well; \citep[e.g.,][]{Croston2004}), and/or it may be related to the gamma-ray emission of 3C 286. 
Assuming a flat powerlaw shape for this component with $\Gamma_{\rm x}$=1.5  
results in an observed, absorbed (3--10 keV) flux of $\sim$3.1 $\times$ 10$^{-15}$ erg\,s$^{-1}$.

\subsection{The origin of the gamma-ray emission of 3C 286}

3C 286 is one of only a small number of radio steep-spectrum sources  which have a Fermi gamma-ray detection. 
A variety of scenarios were discussed in the literature for CSS in general \citep{ODea2021}. In application to 3C 286, \citet{An2017} explored the possibility that the {\em inner} jet is misaligned with the outer jet and is pointing at us and is relativistic, but found that that was not the case, with a viewing angle of the inner jet of 48$^{\rm o}$.  
Here, we discuss several  scenarios for the gamma-ray emission of 3C 286 in turn. 

\paragraph{Binary AGN, binary SMBH, or chance projection of multiple sources} 
Pairs of AGN and binary SMBHs form in the course of galaxy mergers, and many systems have been identified in recent years \citep{KomossaZ2016, deRosa2019}.
If an AGN pair or binary SMBH was present in 3C 286, we could imagine one of them with NLS1–CSS properties, and the other one with blazar characteristics producing the
detected gamma-rays.  It is interesting to note that \citet{Zhang1994} reported the presence of two core-like sources at the center of 3C 286 (C1 in the NE, and C2 in the SW), speculating about these representing two separate nuclei. Their radio spectra and fluxes were found to be very similar. However, follow-up observations revealed SED differences below 8 GHz where the NE source showed an inverted spectrum (based on non-simultaneous data). Therefore, \citet{An2017} interpreted the NE component as the actual core, and the SW component as a knot in the jet. 
Further, we do not find any other indication for the presence of a dual AGN, for instance in form of double-peaked optical emission lines.

One can think of two (or multiple) unrelated sources as well, at different redshifts but 
closely projected 
on the sky. For instance, one source could be a radio-quiet, variable NLS1 galaxy dominating the optical spectrum and UV--X-ray excess, the other a gamma-ray emitting CSS. However, in that case the intervening absorber at $z=0.692$, seen in both the radio and UV, either implies that radio and UV arise from one and the same source, or that both are very close in projection, such that both sightlines pass the intervening absorber. Available Hubble Space Telescope (HST) images did not yet reveal multiple optical sources \citep{deVries1997}.   
We therefore conclude that there is no positive evidence for these scenarios at present.

\paragraph{Intervening galaxy.} 
\citet{Yao2021} 
raised the question, whether the known intervening galaxy at 2.5''  offset from 3C 286 could actually be an AGN and be the counterpart of the gamma-ray emission. However, this was rejected because the galaxy was not detected in the radio band, and is known to have low metallicity. With our deep Chandra observation, we have now checked for any X-ray emission from this galaxy and  do not detect any (Fig. \ref{fig:chandra-ima}). We conclude that it is not the counterpart of  the  Fermi source.

   \begin{figure}[h]
   \centering
  \includegraphics[width=6cm]{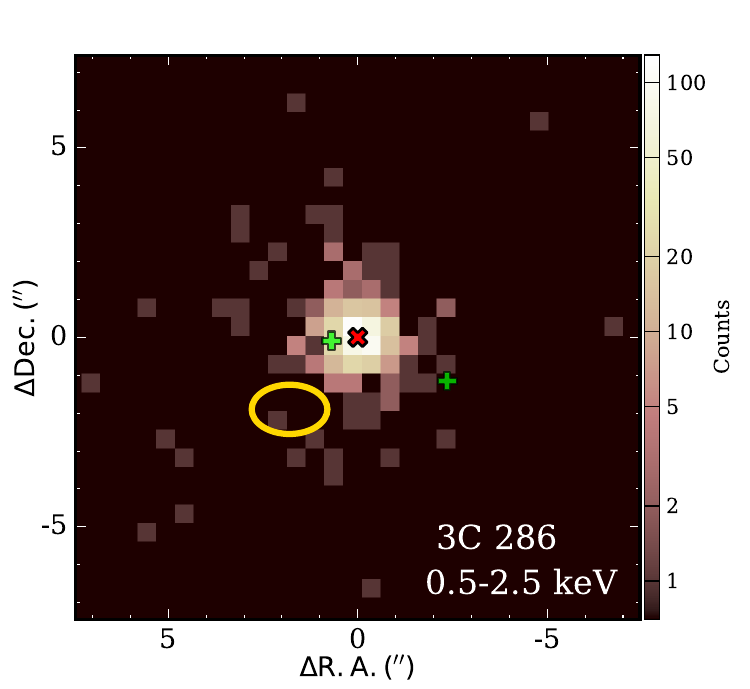}
  \includegraphics[width=6cm]{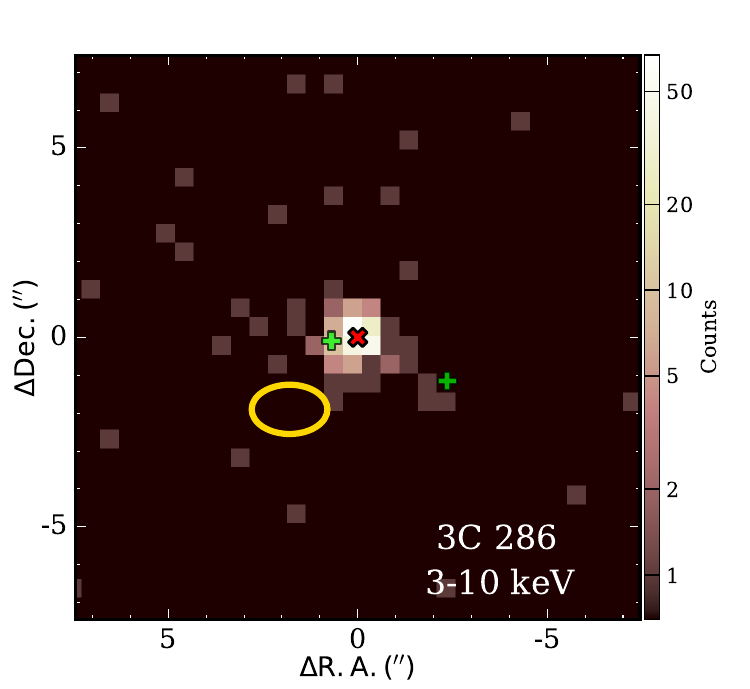}
   \caption{Chandra X-ray images of 3C 286 in the soft {\bf{(0.5--2.5 keV; left)}} and hard {\bf{(3--10 keV; right)}} X-ray band. Locations of several features of interest are marked: the red cross marks the X-ray center of 3C 286. The green plus symbols correspond to the two  radio lobes in the East and South-West, and the yellow ellipse marks the position of the intervening low surface brightness galaxy. 
   }
              \label{fig:chandra-ima}%
    \end{figure}

\paragraph{Other nearby, unrelated, galaxies in the field.}
Given the limited spatial resolution of Fermi, there is a residual small possibility, that the gamma-ray emission does not arise from 3C 286 but rather from an unrelated foreground or background source. However, we find that all other X-ray sources detected (at $>3\sigma$) in the field with Swift and/or Chandra are much fainter in X-rays than 3C 286, by factors $>6-10$. In particular, no 2nd bright source near 3C 286 is detected at the spatial resolution of Chandra.

\paragraph{Microlensing.} Next, we note that we cannot have a situation of microlensing which temporarily intensified any intrinsically much fainter gamma-ray emission,
because Fermi found persistent emission for multiple years.

\paragraph{Emission from lobes.} 
The radio lobes have been considered as potential sites for gamma-ray emission. For instance, it was suggested  that inverse Compton (IC) processes involving local photons   produce high-energy radiation \citep{Stawarz2008, Migliori2014}   if electrons can be efficiently re-accelerated  locally for  instance in hot spots and then injected into the lobes.  
Upcattering of cosmic microwave background (CMB) photons is another mechanism  \citep{Abdo2010, Zhang2020, Yu2024}.
Bremsstrahlung emission in the MeV--GeV regime from shocked plasma created from a cocoon expanding into the ISM was considered, too \cite{Kino2009}. 
In some of these scenarios, we expect enhanced off-nuclear X-ray emission at these sites,  and have searched for  any such emission with Chandra.
The SW lobe is spatially resolved, and 3 photons from its location were detected;  too few to pin down their production mechanism.

It should be noted that some of the models above \citep{Stawarz2008, Kino2009, Migliori2014} are made for young CSS radio sources evolving into a dense medium at one-to-few kpc scales, while 3C 286 is already more evolved (see next paragraph), and no evidence for significant intrinsic absorption along our line-of-sight was found either.

\paragraph{Jet deflection on dense gas.}  Alternatively, it was speculated that individual jet components get occasionally
deflected and then beamed toward the observer as they encounter single dense blobs in the ISM along the path of the jet 
\citep{Bosch-Ramon2012}. 
CSS are generally considered young AGN \citep[review by][]{ODea2021}, their compact jets evolving through the dense ISM. 
Whether this holds for 3C 286 in particular, is less clear, however. Its radio emission is more extended than most CSS (up to a projected size of $\sim$ 20 kpc, 
beyond
the confines of the host galaxy). 
Further, there is no evidence for absorbing
material along our line-of-sight intrinsic to 3C 286  \citep[][our Sect. 6.3]{Grasha2019}. 
The inner region remains spatially unresolved in X-rays, but a highly beamed component would likely produce radio variability, not observed.  

\paragraph{Strong (and variable) external photon field.} 

\citet{Zhang2020} highlighted the fact that gamma-ray emitting CSS, including 3C 286, have on
average higher Eddington ratios than the rest of the CSS population. In particular,
the NLS1 nature of 3C 286 explains the bright soft excess (disk) contribution to the SED
and the high accretion near the Eddington rate, and its reprocessing on BLR and torus scales  may provide an efficient external photon field for the inverse Compton processes. 
Based on the BLR radius--luminosity relation of \citet{Kaspi2005},
and the 5100 \AA~luminosity \citep{Yao2021} of $\lambda\,L_{\lambda}(5100)$=1.6 $\times$ 10$^{45}$ erg\,s$^{-1}$ (estimated from $L(H\beta)$) or up to
$\lambda\,L_{\lambda}(5100)$=4.7 $\times$ 10$^{45}$ erg\,s$^{-1}$ (the directly observed value, assuming the optical emission is dominated by the accretion disk), the BLR radius of 3C 286     is $R=$ 151 -- 318 lt-days.           
 This approach is worthwhile exploring further, carrying out SED  modelling using the most recent X-ray spectral constraints.

\section{Summary and conclusions}

After a short review of the unique MWL properties of 3C 286, we have presented new observations from the radio to X-ray regime. Our main results can be summarized as follows. 

\begin{itemize}

\item{No radio variability is detected. The (2.6--41 GHz) flux density, the polarization percentage, and polarization angle remain constant at Effelsberg telescope resolution when compared to past measurements, confirming the calibrator status of 3C 286.  
}
\item{The X-ray spectrum of 3C 286 is well represented by a soft excess on top of a flat powerlaw.
The longterm X-ray light curve confirms the evidence for X-ray variability by a factor $\sim$2, and spectral fits to Swift observations suggest that it is the soft (disk) component which varies; not unexpected in light of the NLS1 classification of 3C 286. Disk variability will not necessarily affect the inner jet, but could.}
\item{The amount of extinction/absorption along our line-of-sight was rigorously assessed. No reddening intrinsic to 3C 286 is found, and the excess X-ray absorption detected with XMM-Newton \citep{Yao2024} is consistent with the intervening DLA system which has low dust content as well.  The lack of reddening validates the estimate of a high Eddington ratio, the NLS1 classification of 3C 286, and its high radio-loudness index. 3C 286 stands out as one of the radio-loudest NLS1 galaxies known.} 

\item{The Chandra image reveals tentative evidence for hard X-ray emission from the SW lobe. }

\item{Different gamma-ray emission mechanisms were discussed. Several scenarios were found to be unlikely (no positive evidence for a dual AGN or binary SMBH so far, no alternative counterpart to the gamma-ray emission detected in X-rays, no evidence that 3C 286 is a particularly {\em young} CSS with its jet still making its way out of the galaxy for the first time). It is speculated that the NLS1 character of the source and its high Eddington ratio, and efficient reprocessing of the disk photons in the BLR and torus, lead to efficient IC processes contributing to the observed gamma-ray emission.}
\end{itemize}

A number of follow-up observations of 3C 286 suggest themselves. 
Deeper XMM-Newton observations will provide tight constraints on spectral variability, and a deeper Chandra observation is needed to confirm the presence of off-nuclear hard X-ray emission
from the lobe(s). 
It will require the next generation of X-ray spectroscopic missions to detect individual metals in the intervening absorber at $z$=0.692.  
Continuing Swift monitoring will allow to quantify the frequency and amplitude of MWL variability. 
JWST imaging will reveal the properties of the intervening low-surface brightness galaxy which is the likely host of the DLA system, and will detect the host galaxy of 3C 286 itself for the first time with high significance (so far only marginally detected with HST; \citep{deVries1997}). VLBI at highest resolution will determine the compactness of the core so far always resolved in radio observations, and reveal variability of the inner jet if present.  

3C 286, among the first quasars identified spectroscopically in the 1960s, and among the first to be imaged with VLBI, continues to be an important cornerstone system for our understanding of the radio-loudest quasars known, the drivers behind NLS1 galaxies and their link with CSS sources, and the mechanism of gamma-ray emission in misaligned sources.  

\section{Appendix}
\subsection{Flux density measurements}

Here we report the flux density and polarization measurements of 3C 286 obtained with Swift (Tab. \ref{tab:magnitudes-swift}) and the Effelsberg telescope (Tab. \ref{tab:fluxdensities-radio}).

\begin{table}[H]
\caption{Swift UVOT magnitudes of  3C 286, corrected for Galactic extinction. Uncertainties are $\pm{0.1}$ mag in the optical filters (V, B, U), and 
$\pm{0.05}$ mag in the UV filters in 2020 and 2024. They are larger in 2023 because of the satellite drift motion, with 
$\pm{0.3}$ mag in the optical filters, and 
$\pm{0.1}$ mag in the UV filters. 
}
	\label{tab:magnitudes-swift}
	\begin{tabular}{lllllll}
		\toprule
		date &  V & B & U & W1 &  M & W2 \\
	\midrule        
2020-08-16 &  17.30 &17.67  &16.33 &15.99  &15.79  &16.00   \\
2020-08-21 &  17.23 &17.59  &16.29 &16.01 &15.81 &16.00 \\
2023-11-04 &  17.42 &17.45  &16.31 &16.14 &15.93 &16.14 \\
2023-11-14 &  17.25 &17.77  &16.35 &16.19 &15.92 &16.16 \\
2023-11-29  &  ... &17.55 &16.58 &16.17 &15.89 &16.11 \\
2023-12-03 & 17.35 &17.30 &16.37 &16.08 &16.02 &16.08 \\
2024-04-18 & 17.34 &17.77 &16.55 &16.19 &16.07 &16.12 \\
		\bottomrule
	\end{tabular}
\end{table}  

\begin{table}[H]
\caption{Radio flux density $S$ and polarization percentage $p$ of 3C 286 measured with the Effelsberg 100m telescope on 2023 November 3 and 16. }
	\label{tab:fluxdensities-radio}
	\begin{tabular}{lllll}
		\toprule
		 $\nu_{\rm c}$ & $S$  & & $p$ &  \\
        GHz & Jy & & & \\
        & 2023-11-03  & 2023-11-16 & 2023-11-03  & 2023-11-16   \\
		\toprule
2.595  & 10.59$\pm{0.41}$ & 10.65$\pm{0.08}$ & 10.9$\pm{0.6}$ & 10.8$\pm{0.4}$   \\
2.95  &   9.76$\pm{0.28}$ & 9.97$\pm{0.21}$ & 11.4$\pm{0.5}$ & 11.0$\pm{0.5}$  \\
4.85  &   7.29$\pm{0.22}$ & 7.52$\pm{0.03}$ & 11.5$\pm{1.2}$ &  11.0$\pm{0.4}$ \\
10.45  &  4.48$\pm{0.18}$ & 4.44$\pm{0.08}$ & 11.3$\pm{1.0}$ & 11.6$\pm{0.7}$  \\
14.25  &  3.49$\pm{0.17}$ & 3.43$\pm{0.19}$ & 12.6$\pm{0.8}$ & 13.1$\pm{0.9}$  \\
19.25  &  2.93$\pm{0.19}$ & 2.83$\pm{0.10}$ &          &   \\
22.85  &  2.46$\pm{0.22}$ & 2.43$\pm{0.11}$  &          &   \\
24.75  &  2.33$\pm{0.29}$ & 2.25$\pm{0.08}$ &          &   \\
36.25  &                  & 1.68$\pm{0.16}$ & &   \\    
41.25  &                   & 1.55$\pm{0.10}$  & &  \\ 
		\toprule
	\end{tabular}
\end{table}

\subsection{Other X-ray sources detected in the field with Swift}

The individual Swift X-ray images between 2020 and 2023 were co-added and a source detection was performed on the merged data set. 
All other X-ray sources are much fainter than 3C 286 itself. 
In Tab. \ref{tab:source-detection}, all sources which are detected above 3$\sigma$ are listed. With NED, optical-IR counterparts were searched within a region of 5'' radius. Two sources have WISE \citep{Wright2010} counterparts, and one coincides with a high-redshift SDSS quasar candidate.  

\begin{table}[H]
\caption{X-ray sources detected with Swift at $>3\sigma$ in the merged 2020 -- 2023 data set,  their average count rates CR in the 0.3--10 keV band, their detection significance $\sigma$, the nearest candidate optical--IR counterpart within 5'' including photometric redshifts if available, and the separation $d$ of the nearest counterpart in arcminutes. 
} 	
 \label{tab:source-detection}
	\begin{tabular}{lcclcll}
		\toprule
source & RA-J2000 & Decl-J2000 & CR  & $\sigma$ & counterpart & $d$ \\
         & ($h,m,s$) & ($^o$,','') & 10$^{-3}$ cts/s &  &  &  \\
      (1)  & (2) & (3) & (4) & (5) & (6) & (7) \\ 
		\toprule
   X1 & 13 31 08.4 & +30 30 36.0 & 15.7$\pm{1.4}$ &  11.1 &  3C 286 & 0.056' \\
   X2 & 13 30 56.0 & +30 40 23.7  &  2.53$\pm{0.60}$ & 4.2 & WISE source & 0.057' \\ 
    X3 & 13 30 33.1 & +30 33 29.2 &   1.90$\pm{0.51}$   &  3.7 & SDSS QSO candidate & 0.069'\\
      & & & & & $z$=1.745 & \\
      X4 & 13 30 57.2 & +30 37 46.2 &   1.50$\pm{0.46}$ & 3.2  & WISE source & 0.069' \\
      X5  & 13 30 30.4 & +30 33 54.4 &   1.53$\pm{0.47}$ & 3.3 & -- & \\
      X6  & 13 31 16.4 & +30 35 52.1  &  1.50$\pm{0.47}$ & 3.2 & --  & \\   
    		\toprule
	\end{tabular}
\end{table}

\vspace{6pt} 




\authorcontributions{
Conceptualization and methodology, SK; data analysis, all authors; writing—original draft preparation, S.K.; writing—review and editing, all authors. 
All authors have read and agreed to the published version of the manuscript.
}

}


\dataavailability{
All data are available upon reasonable request. The Swift data we proposed are available in the Swift archive at \url{https://swift.gsfc.nasa.gov/archive/}. 
} 



\acknowledgments{
 We would like to thank the Swift team for carrying out the observations we proposed. 
 SK acknowledges the hospitality of NAOC Beijing in October 2023.  
 SY acknowledges support by an Alexander von Humboldt Foundation Fellowship between 2020 and 2022, when this project was started. 
We would like to thank our referee and T. Krichbaum for their useful comments on the manuscript.
 This work is partly based on data obtained with the 100\,m telescope of the Max-Planck-Institut f\"ur Radioastronomie at Effelsberg.   
This work made use of data supplied by the UK Swift Science Data Centre at the University of Leicester \citep{Evans2007}. 
This research has made use of the XRT Data Analysis Software (XRTDAS) developed under the responsibility of the ASI Science Data Center (ASDC), Italy.
This research has made use of data obtained with the Chandra satellite, and software provided by the Chandra X-ray Center (CXC).
This work has also made use of the NASA Astrophysics Data System Abstract Service (ADS), and the NASA/IPAC Extragalactic Database (NED) which is operated by the Jet Propulsion Laboratory, California Institute of Technology, under contract with the National Aeronautics and Space Administration.
 }

\conflictsofinterest{The authors declare no conflict of interest.
} 





\begin{adjustwidth}{-\extralength}{0cm}

\reftitle{References}

\PublishersNote{}
\end{adjustwidth}
\end{document}